\documentclass[10pt,letterpaper]{article}
\usepackage[top=1.0in,left=2.5in,footskip=0.75in]{geometry}

\usepackage{amsmath,amssymb}
\usepackage{bm}

\usepackage{changepage}

\usepackage{textcomp,marvosym}

\usepackage{cite}

\usepackage{siunitx}

\usepackage{nameref,hyperref}

\usepackage[right]{lineno}
\usepackage{float}
\usepackage[dvipsnames]{xcolor}

\usepackage[expansion=false,nopatch=eqnum]{microtype}
\DisableLigatures[f]{encoding = *, family = * }

\usepackage[table]{xcolor}

\usepackage{array}

\newcolumntype{+}{!{\vrule width 2pt}}

\newlength\savedwidth

\usepackage{setspace} 
\usepackage[aboveskip=1pt,labelfont=bf,labelsep=period,justification=raggedright,singlelinecheck=off]{caption}

\makeatletter
\renewcommand{\@biblabel}[1]{\quad#1.}
\makeatother

\usepackage{lastpage,fancyhdr,graphicx}
\usepackage{epstopdf}
\fancyheadoffset[L]{2.25in}
\fancyfootoffset[L]{2.25in}
\begin{document}
\vspace*{0.2in}

\begin{flushleft}
{\Large
\textbf\newline{Fully-Automated Multi-Modal Registration of Timelapse Zebrafish OCT+LS Images} 
}
\newline
\\
Mahshid Babeli Nia\textsuperscript{1},
Leah A. Lewis\textsuperscript{2},
Mohammad Mobarak Karim\textsuperscript{2},
Manmohan Singh\textsuperscript{2},
Arne C. Lekven\textsuperscript{3},
Guoning Chen\textsuperscript{4},
Kirill V. Larin\textsuperscript{2},
David Mayerich\textsuperscript{1*}
\\
\bigskip
\textbf{1} University of Houston, Department of Electrical and Computer Engineering, Houston, TX, 77204, USA
\\
\textbf{2} University of Houston, Department of Biomedical Engineering, Houston, TX, 77204, USA
\\
\textbf{3} University of Houston, Department of Biology and Biochemistry, Houston, TX, 77204, USA
\\
\textbf{4} University of Houston, Department of Computer Science, Houston, TX, 77204, USA
\\

\bigskip

\end{flushleft}

\section*{Abstract}
Optical coherence tomography (OCT) and light sheet (LS) fluorescence microscopy provide complementary views of tissue morphology and molecular expression, offering a powerful approach for studying embryonic development. However, longitudinal integration of OCT+LS images is challenging because the modalities differ substantially in contrast, spatial resolution, field of view, and visible anatomical features, while embryo growth and movement introduce additional misalignment over time. These challenges stem from both the scale and multimodal nature of the OCT and LS image stacks. We present a fully automated three-dimensional rigid registration framework for aligning OCT and LS volumes of developing zebrafish embryos acquired from 21 to 48 hours post fertilization (HPF). The method does not require pre-existing segmentations or cross-modal feature correspondences, which are often unreliable because OCT and LS capture fundamentally different tissue properties.
Registration performance was evaluated across developmental time points using mean target registration error (mTRE) and compared across multiple similarity metrics and optimization methods. Normalized cross-correlation combined with particle swarm optimization produced the most accurate and consistent registration, achieving the lowest mTRE at every evaluated time point compared with the alternative methods. The registration establishes spatial correspondence between OCT tissue morphology and LS fluorescence signals throughout development, enabling integrated longitudinal analysis of structural and molecular changes in living embryos.


\section{Introduction}

Studying molecular pathways in embryo development requires imaging three dimensional morphology with co-registered molecular distributions \cite{shah2019multi}. Two promising methods include optical coherence tomography (OCT) \cite{lichtenegger2022optical} and light sheet (LS) fluorescence microscopy \cite{keller2008reconstruction,vladimirov2014light}. OCT provides tissue morphology based on differences in refractive index, while LS reveals the distribution of specific molecular markers such as proteins \cite{lawson2002vivo}. Since neither modality alone can deliver both structural and molecular information simultaneously, multimodal studies combining the two could be particularly valuable for investigating specific molecular pathways, provided the images are spatially and temporally co-registered.

\begin{figure}[H]
    \centering
    \includegraphics[width=1\linewidth]{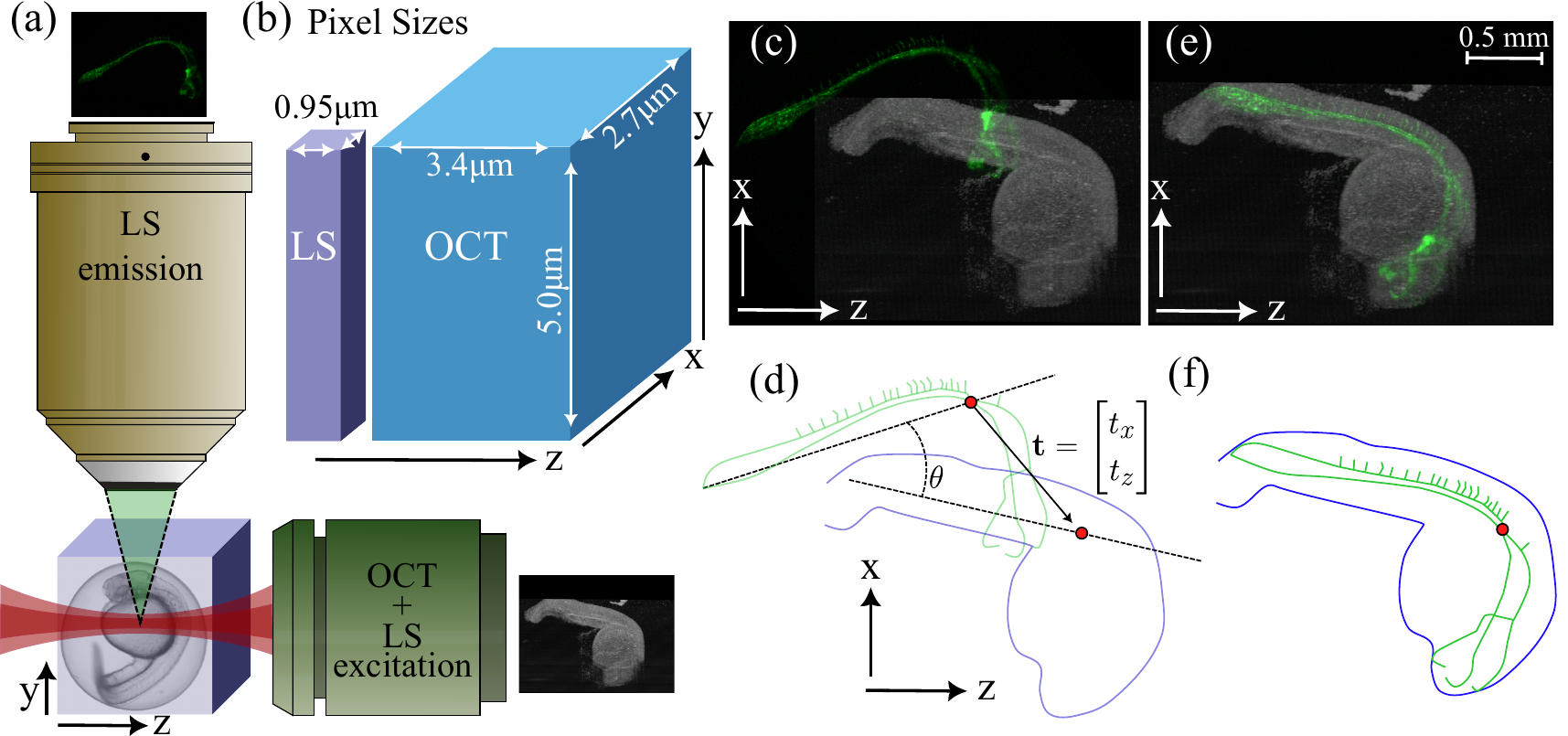}
	\caption{OCT+LS imaging applies constraints to optimized registration. (a) A schematic of the sample region of the combined OCT+LS imaging system illustrates the imaging geometry of the two modalities. The OCT imaging mode shares an excitation path with the LS mode, ensuring that the images are closely aligned along the $X$ axis. Note that the $Z$ axis is defined by the axial direction of the OCT mode, therefore the LS image is collected in the $YZ$ plane. (b) Both modalities enforce constraints on optical resolution, making the pixel sizes both anisotropic and inconsistent between modalities. The sample rate along the $Y$ axis matches between modalities, and is constrained by the thickness of the LS light sheet. (c-d) After compensating for pixel size, OCT+LS image pairs are primarily misaligned in the $YZ$ plane. This rigid transformation consists of a translation $\mathbf{t}$ and rotation $\theta$ that must be calculated during optimization. (e-f) After rigid registration, both modalities are aligned within a single time point.}
    \label{fig:octls_imaging}
\end{figure}

OCT+LS imaging \cite{khajavi2021multimodal} combines OCT and LS into a single instrument to simultaneously provide microstructure and molecular information from an \textit{in vivo} sample. Paired OCT and LS volumes are acquired across multiple time steps using a co-planar OCT+LS imaging system (Figure \ref{fig:octls_imaging}). However, aligning these images across modalities and time remains challenging because LS and OCT differ in contrast, field of view, resolution, and spatial features. Although the $y$-axis planes are aligned within each OCT+LS pair, they are not necessarily aligned across different time steps as the embryo grows and moves. 

OCT+LS alignment between modalities is particularly challenging due to: (1) a lack of features shared between modalities, (2) large pixel counts in volumetric image stacks that make comparison time consuming, (3) a large number of image pairs and time steps, and (3) artifacts such as speckle noise, signal attenuation, and shadowing that confound similarity metrics. Since the goal of OCT+LS is the acquisition of volumetric image sequences demonstrating complex phenotypic changes, adequately constraining the optimization methods after acquisition is impractical.

In this paper, we present a fully-automated rigid registration framework for longitudinal OCT+LS imaging of zebrafish embryos. The proposed method aligns paired OCT and LS volumes across multiple developmental time points without requiring pre-existing segmentations or dense cross-modal correspondences. By estimating rigid transformations between image volumes, the framework enables consistent tracking of structural and molecular changes during embryonic development.

\section{Background}

Image registration is the process of aligning images by estimating a spatial transformation that maps spatial locations in both images to a common shared space. Transformation models are classified as rigid, affine, or deformable according to the types of spatial transforms that are allowed. Rigid transformations preserve distances and angles using translations and rotations, while affine transformations additionally allow scaling and shear. Deformable transformations capture complex local deformations \cite{reithmeir2025model}, which are often needed to align samples that are morphologically different.

\subsection{Similarity metrics}
\label{subsec:metrics}
The similarity metric quantifies how well two images match. In the case of registration, this metric aims to provide an objective assessment of a given spatial transformation. The sum of squared differences (SSD) is computationally efficient but assumes similar image intensities and is most suitable for monomodal registration \cite{andersson2007non}. Normalized cross-correlation (NCC) is more robust to global brightness and contrast differences\cite{evangelidis2008parametric}. Because NCC measures correlation over the whole image, it may be less suitable for non-rigid registration, where small regions can undergo local changes. Local normalized cross-correlation (LNCC) \cite{balakrishnan2019voxelmorph} addresses this limitation by evaluating correlation within local neighborhoods. Normalized mutual information (NMI) measures the statistical dependence between the intensity distributions of images \cite{viola1997alignment}, but is susceptible to noise that reduces the accuracy of the estimated statistics. Changes in anatomy or image contrast can also alter the statistical relationship between images. The modality-independent neighborhood descriptor (MIND) compares local self-similarity patterns rather than raw intensities, making it more robust to contrast differences \cite{heinrich2012mind}. However, MIND is computationally more expensive than similar intensity-based metrics, and its performance depends on the choice of neighborhood size and descriptor parameters that are difficult to optimize.

\subsection{Optimization methods}

The optimization method estimates transformation parameters by maximizing or minimizing the similarity metric. Gradient-based methods \cite{lecun2002efficient}, such as gradient descent \cite{ruder2016overview} or Adam \cite{kingma2014adam}, update the transformation based on the local partial derivatives of the similarity metric. They are computationally efficient but can easily get caught in local optima instead of finding the best overall solution. Population-based methods \cite{rahimi2023review}, such as Evolution strategies (ES) \cite{salimans2017evolution} or particle swarm optimization (PSO) \cite{kennedy1995particle, shami2022particle}, evaluate multiple candidate solutions across the search space. They are less likely to become trapped in local optima but generally require greater computational cost due to increased sampling.

\subsection{Multimodal image registration}
Rigid multimodal registration is frequently used in remote sensing when locations are imaged using different sensors or lighting conditions. These applications frequently rely on feature-based methods \cite{zhang2025multi, velesaca2024multimodal} that match keypoints between images and estimate the geometric transformations that would align them. These predominantly include the scale-invariant feature transform (SIFT) \cite{lowe2004distinctive} and oriented FAST and rotated BRIEF (ORB) \cite{rublee2011orb, moghimi2021comparison}.

SIFT extracts keypoints and descriptors based on local characteristics such as intensity and texture. The algorithm then matches features between images by comparing descriptor vectors. The key challenge in multimodal alignment is that these local descriptors are inconsistent between modalities, making it difficult to match features between images. 

Several modern approaches attempt to harmonize features across modalities. Phase congruency-based methods \cite{ye2024fast, li2019rift} use multiscale and multi-orientation log-Gabor filters to extract structural information that is more consistent across modalities. Filter responses are combined across scales and orientations to identify stable structural features that have consistent phase responses. The resulting phase-congruency maps highlight structural features, such as corners and edges, for multimodal registration. Cross-modal SIFT (CSIFT) \cite{lian2025csift} improves feature descriptors using circular Gaussian-weighted oriented gradient features (CGOGF), which assign higher weights to orientation directions closer to each pixel’s gradient direction and lower weights to directions with greater orientation differences, preserving more structural information across modalities. Other method improves image matching by introducing feature position, scale, and orientation constraints into the matching process. These additional geometric constraints reduce false matches and improve registration robustness \cite{zhang2025optical}. 

\begin{figure}[H]
    \centering
    \includegraphics[width=0.85\linewidth]{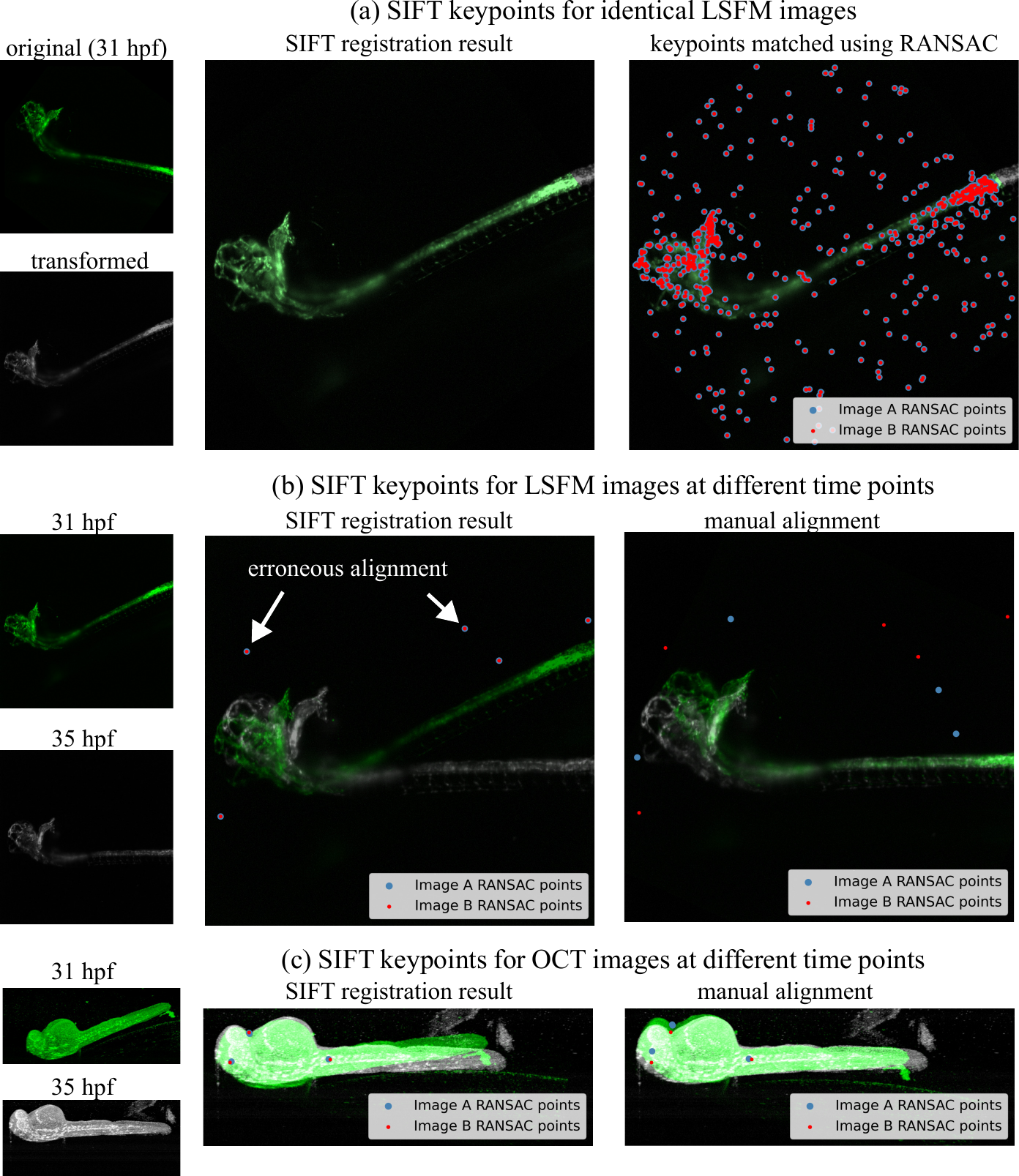}
	\caption{Challenges in identifying common keypoints between time points in evolving images. (a) A rigid transform is applied to a static LS image of a zebrafish acquired at 31 hours post fertilization (HPF). SIFT (filtered by RANSAC) identifies dozens of common keypoints between both images, facilitating alignment. (b) LS images acquired at different timepoints contain very few matching features. The 5 corresponding points identified are erroneous, resulting in poor registration. (c) The same analysis is shown for OCT images at different time points. While OCT images of the zebrafish are more consistent over time, the detected keypoints still provide poor alignment compared to the ground truth.}
    \label{fig:feature_problems}
\end{figure}

Identifying reliable cross-modality features is particularly challenging in OCT+LS because the modalities capture distinctly different components of the sample \textbf{by design}: OCT provides global tissue structure, while LS highlights individual proteins tied to internal structures invisible to OCT. The goal of OCT+LS is to leverage multiple channels that capture \textit{distinct} tissue components unique across channels and modalities. Therefore, the same anatomical location can appear very different, making feature matching unreliable for image registration (Figure \ref{fig:feature_problems}).

\subsection{Registration in biomedical imaging}

The primary clinical application for multimodal registration is alignment of patient computed tomography (CT) and magnetic resonance imaging (MRI) data. Feature-based registration can improve multimodal alignment by emphasizing structural information prior to registration. For example, one MRI-CT registration method combined intensity-based registration with feature extraction to refine the initial alignment \cite{arora2024adaptive}. Another study designed feature descriptors using either pretrained segmentation models or MIND descriptors to compare extracted features using SSD \cite{siebert2024convexadam}. More recently, a CT-MRI registration method \cite{chen2025dual} performed coarse alignment after enhancing edge structures using phase congruency and Sobel edge detection, then applied fine alignment using a modified mutual information metric and optimization. Similarly, a recent unsupervised multimodal registration method extracted robust local features for histology and mass spectrometry (MS) tissue images, including H\&E sections and mouse liver MS images \cite{shojaei2025novel}. The method used multi-scale features from both images and used hierarchical average pooling to describe regions with high local intensity variation, and matched these features to estimate an affine transformation.

The Learn2Reg challenge provides a standardized benchmark for medical image registration \cite{hering2022learn2reg}. Its goal is to promote the development of generalizable registration methods applicable to different anatomies (ex. brain, abdomen, lung) across modalities (ex. CT, MRI, ultrasound). The challenge includes several tasks, such as ReMIND2Reg \cite{dorent2025brain, juvekar2023brain} for MRI/ultrasound alignment in brain surgery, and LLUMIR \cite{chen2026beyond} for large-scale unsupervised inter-subject brain MRI registration. Deep learning methods are currently the most popular, and the challenge provides data sets for training and evaluation with images from different scanners, patient groups, and imaging protocols. 

Over the duration of this challenge, successful methods have moved toward deep learning and coarse-to-fine deformations. These methods predict dense displacement fields using CNN-based architectures. Early work like SITReg improves stability by making the deformation symmetric and consistent \cite{honkamaa2023sitreg}. More recent work uses intensity randomization and structural similarity to learn multi-contrast brain MRI registration without ground-truth deformation fields \cite{liu2026zero}. Recent advances such as CLEM-Reg \cite{krentzel2025clem} have successfully aligned volumetric images from very distinct modalities, such as fluorescence and electron microscopy. However, these techniques rely on a pre-existing segmentation that separates anatomy into meaningful regions used to guide alignment. In whole-body PET-CT registration \cite{jonsson2022image}, skeletal regions were first aligned separately using affine transformations which then initialized deformable registration. Subsequent work on CT-MRI registration \cite{walluscheck2023mr} used segmented brain structures to perform registration with normalized gradient fields. These approaches work best when both modalities show anatomical structures that can be segmented and used as common guidance for alignment. However, in OCT+LS imaging, such segmentations are not readily available. OCT captures global tissue morphology while LS highlights specific fluorescent markers, producing fundamentally different image appearances that make consistent, reliable segmentation across modalities difficult to obtain. As a result, segmentation-guided registration methods cannot be directly applied to OCT+LS imaging.

\section{Materials and methods}

\subsection{Tissue preparation}
This study used transgenic zebrafish that express green fluorescent protein (GFP) in endothelial cells (primarily blood vessels).  Tg(kdrl:eGFP) zebrafish embryos were collected immediately after fertilization and incubated in their culture dishes containing E3 media (5 mM NaCl, 0.17 mM KCl, 0.33 mM $CaCl_2$, and 0.33 mM $MgSO_4$ in distilled water) at 29°C. At 20 hours post fertilization (HPF), the health of the embryos was checked with a brightfield microscope (Stemi 508, Carl Zeiss AG, Germany). Healthy embryos were then examined with an epifluorescence microscope (Axioscope 5, Carl Zeiss AG) to confirm fluorescence expression.

Zebrafish were anesthetized using tricane during imaging. A 20X stock solution of tricaine (MS-222, TMS, tricaine methanesulfonate) was diluted to 1X before use. A single fluorescent zebrafish embryo was selected for imaging and anesthetized to prevent movement during imaging by being transferred to a culture dish containing 0.02\% tricaine in E3 media. The chorion was manually removed from the embryo to enable clear visualization while imaging. A concentration of 0.3\% low-melting point agarose was used to maintain the sample positioning during imaging, while still allowing growth. Tricaine was mixed into the warm agarose solution to reach a final working concentration of 0.02\%. The zebrafish embryo, along with the tricaine-agarose mixture was drawn up and pipetted into a glass cuvette, which was mounted on the sample stage for imaging. Images of the same embryo were acquired every 30 minutes from 21 HPF – 48 HPF. Additional tricaine/E3 was pipetted into the cuvette throughout the experiment to maintain sample anesthesia.

\subsection{Image acquisition}
Images were acquired using a custom optical coherence tomography (OCT) and light sheet (LS) multimodal system (OCT+LS) \cite{khajavi2021multimodal}. LS excitation was produced using a 488 nm laser focused through a low numerical aperture objective (iChrome MLE, Toptica Photonics Inc., Farmington, NY). A scanning mirror was used to sweep the beam across the sample to produce a light sheet with a thickness of $\sim$11~$\mu$m. Fluorescence emission was collected using a (NA = 0.8) 16$\times$ water-dipping objective (N16LWD-PF, Nikon Corporation) and passed through a 520 $\pm$ 5 nm bandpass filter to a digital camera (C11440-22CU, Hamamatsu). This provided a transverse resolution of 2.1 µm and a field of view of 1.81 mm $\times$ 1.81 mm. 

OCT used a swept source laser with a central wavelength of 1051 nm, 109 nm bandwidth, 100 kHz sweep rate, and incident power at the sample 6 mW. The system was based on a Michelson interferometer with a sensitivity of 100.1 dB and a 6.9 dB sensitivity roll-off over 4.1 mm. The axial and the lateral resolutions of OCT were 7.6 µm and 14.9 µm, respectively. The OCT beam passed through the polarization beam splitter, with one beam directed to the reference mirror and the other combined with the LS beam. The LS and OCT beams were aligned such that the excitation sheet and OCT beam lay in the same plane, allowing co-registration. 

\subsection{OCT+LS imaging constraints}
We define the shared OCT+LS coordinate system such that the $z$ axis corresponds to the excitation direction for both modalities (Figure \ref{fig:octls_imaging}a). However, the collected OCT and LS images have different sampling rates (voxel sizes) along each axis due to physical constraints of the modalities. Since the OCT and LS excitation beams share an optical path, the $xz$ plane for both modalities is closely aligned and they share a sample rate along the $y$ axis (\SI{5}{\micro\meter} per voxel). The $xz$ plane coincides with the focal plane for the LS emission objective; therefore the $z$ and $x$ sampling rates for LS are isotropic (\SI{0.95}{\micro\meter}). The $z$ and $x$ sampling rates for OCT depend on the depth sampling of each A-line generated by the swept-source interferometric acquisition and the galvanometer scan spacing across the imaging plane, respectively (Figure \ref{fig:octls_imaging}b). 

To register both modalities, we rely on constrained optimization. We first rescaled both images using their known voxel sizes. Because the OCT and LS paths are closely aligned, it is within the $xz$ plane that major misalignments can occur. These include a translation $\mathbf{t}$ within the $xz$ plane and a rotation $\theta$ along the $y$ axis (Figure \ref{fig:octls_imaging}d). Although the OCT and LS excitation beams are designed to be co-planar, in practice there is always some unavoidable alignment error. To account for this, we allowed some constrained registration along the $y$ axis, as well as constrained rotations along the $z$ and $x$ axes.

\subsection{Registration metrics}
\begin{figure}[H]
    \centering
    \includegraphics[width=1\linewidth]{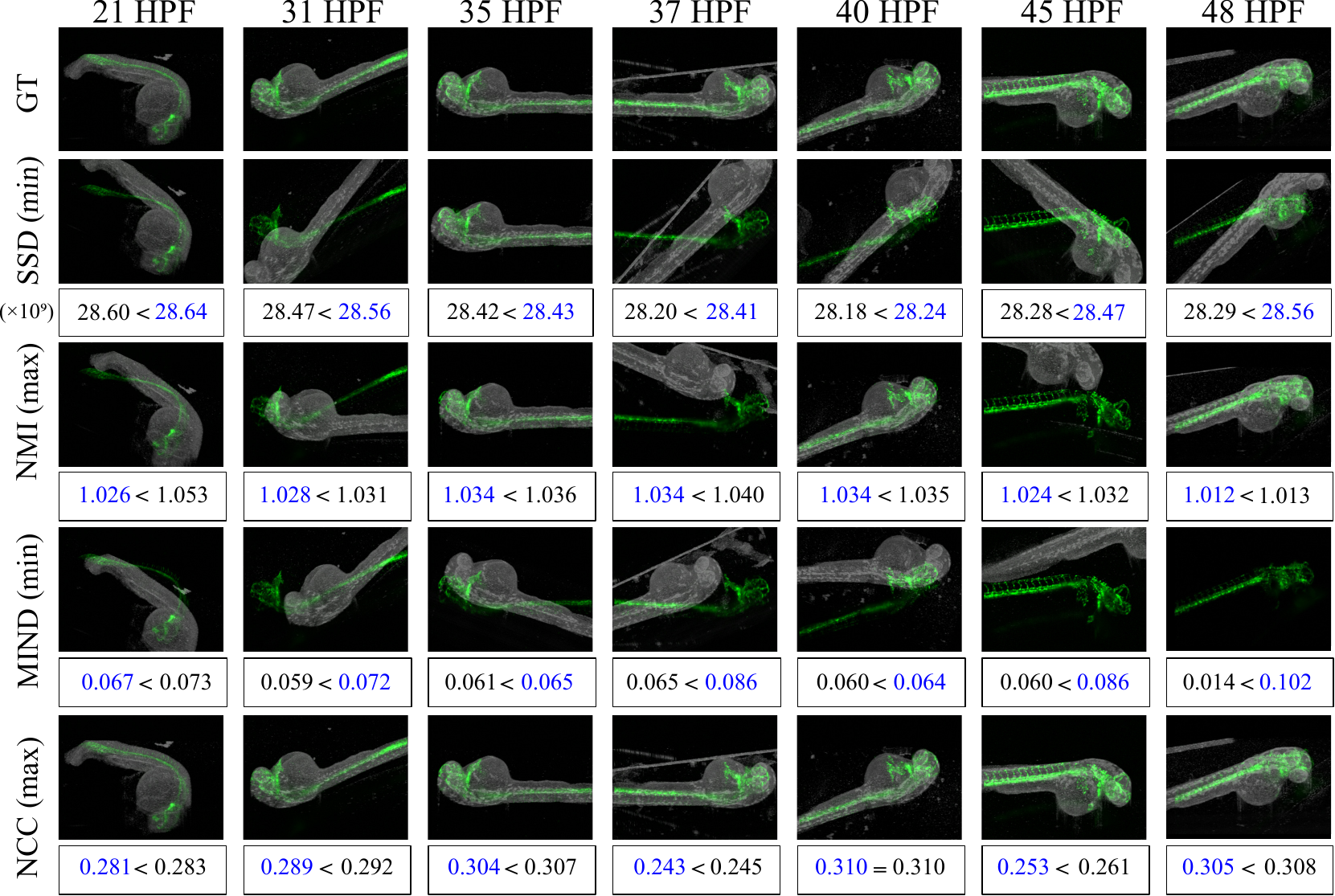}
    \caption{\textbf{Unconstrained} registration results for various metrics applied to multimodal alignment of OCT to LS images. A series of images were collected from a developing zebrafish embryo across time points: 21, 31, 35, 37, 40, 45, and 48 HPF. Images are shown as maximum intensity projections along the Y-axis. The top row shows the manually-aligned ground truth (GT) for each time point. Following rows show the alignment result with the LS image kept constant. The final aligned metric result is shown, along with the corresponding result from the GT. Poorly-performing metrics result in optimal values that do not correspond with an aligned image (blue). NCC exhibited the most consistent performance, where higher NCC values correspond to optimal image alignment.}
    \label{fig:metric_comparison}
\end{figure}

\begin{figure}[H]
    \centering
    \includegraphics[width=1\linewidth]{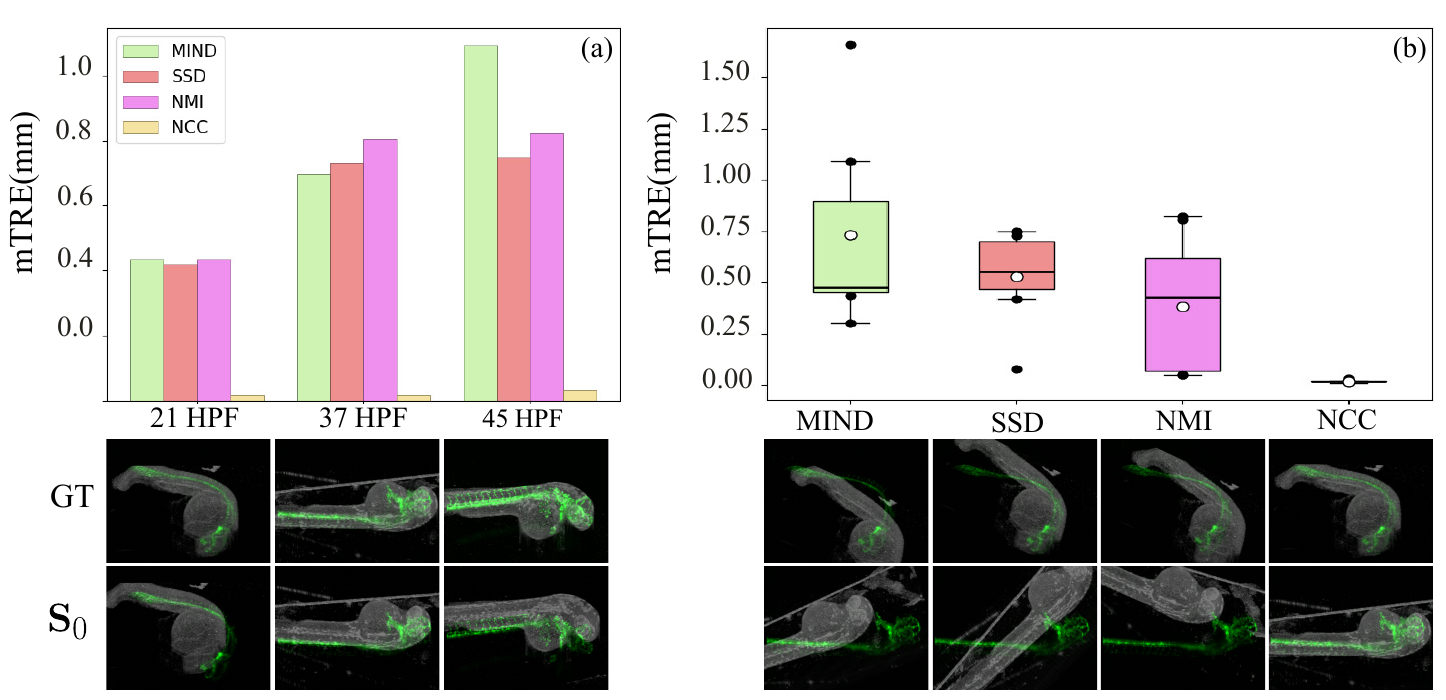}
    \caption{Comparison of similarity metrics using mean target registration error (mTRE) based on \textbf{unconstrained} registration. (a) mTRE values for MIND, SSD, NMI, and NCC at three representative time points (21, 37, and 45 HPF), with the corresponding GT and initial alignments shown below. (b) Distribution of mTRE values for each metric across all seven time points, with two representative registration results shown below for each metric.}
    \label{fig:metric_plot}
\end{figure}

\begin{figure}[H]
    \centering
    \includegraphics[width=1\linewidth]{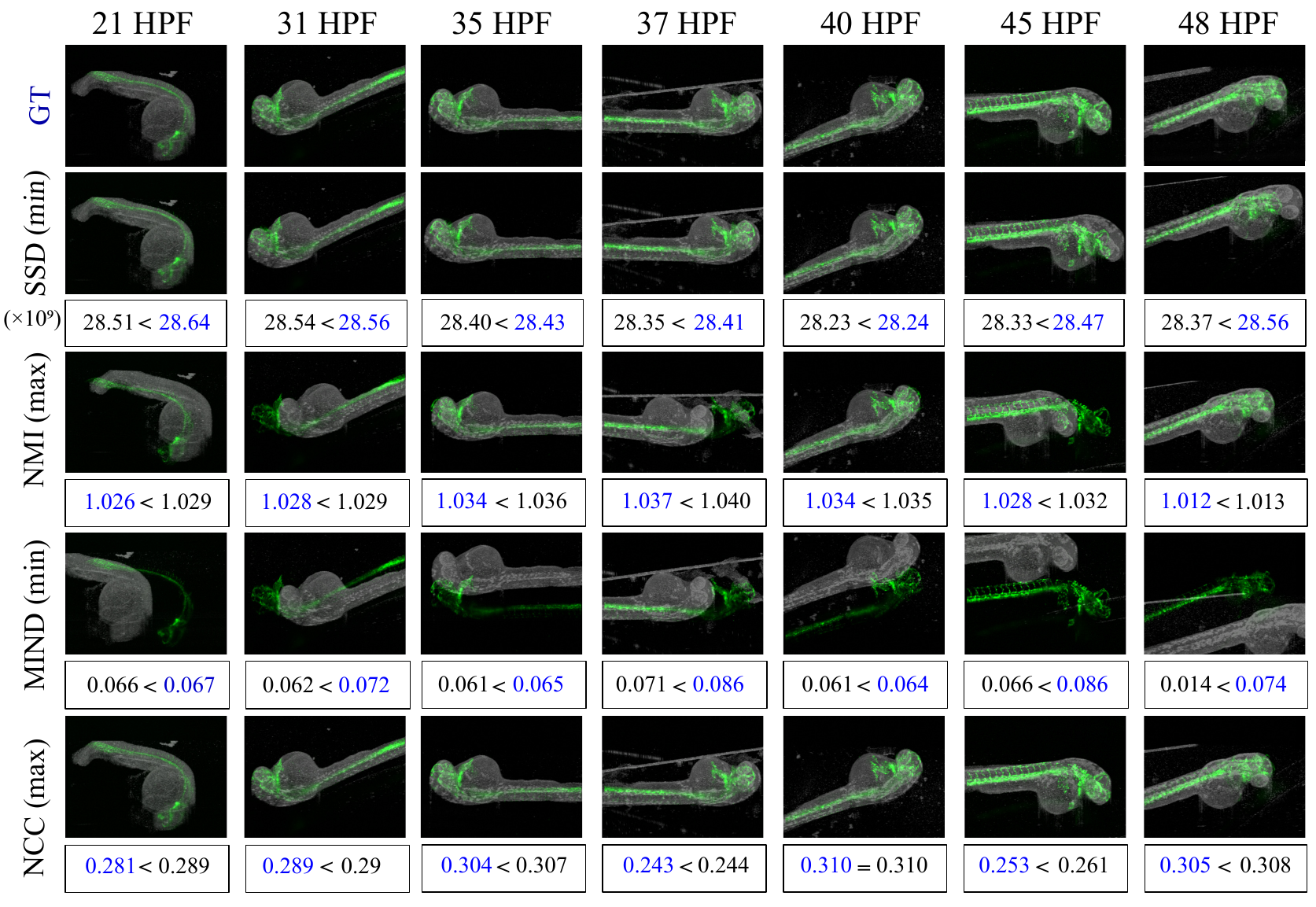}
    \caption{\textbf{Constrained} registration results for the same samples shown in Figure \ref{fig:metric_comparison}.}
    \label{fig:Constrained_metrics}
\end{figure}

\begin{figure}
    \centering
    \includegraphics[width=1\linewidth]{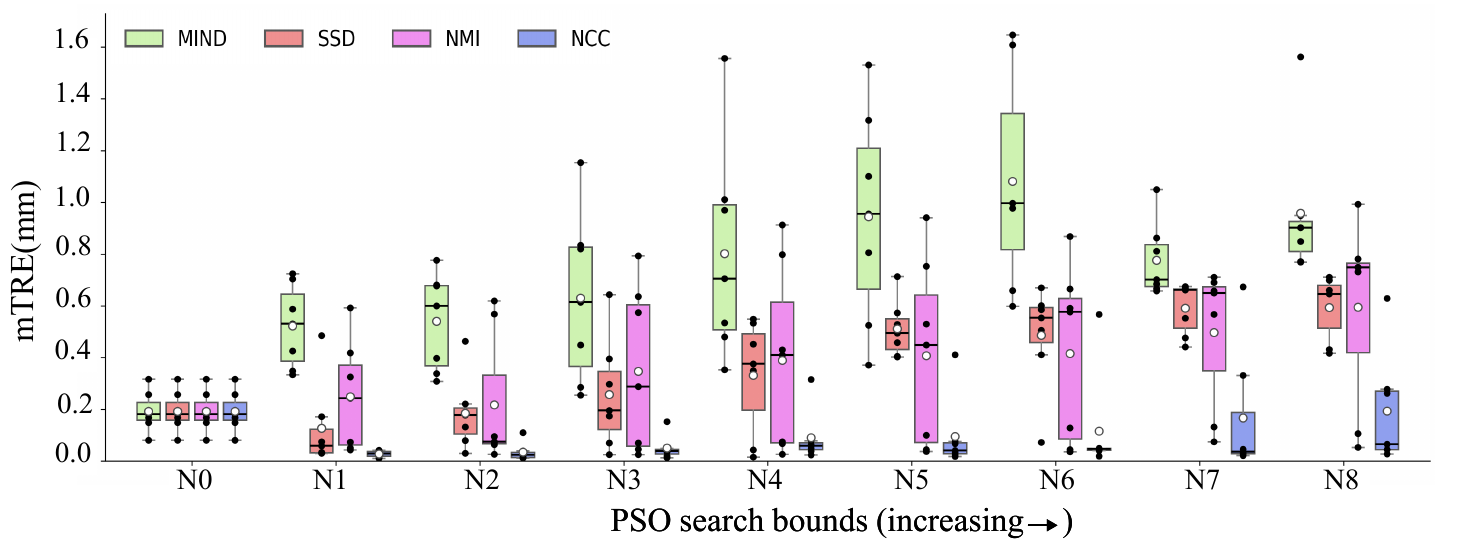}
    \caption{Mean registration error (mTRE) across timepoints for NCC, SSD, NMI, and MIND under increasing PSO search bounds (N1–N8). N0 represents the initial alignment before optimization. Boxes represent the interquartile range (25th–75th percentiles), with horizontal lines indicating medians and white circles indicating means. Whiskers represent the range excluding outliers, while black dots denote individual measurements at each time point. NCC achieves the lowest mean mTRE across all evaluated PSO search ranges.}
    \label{fig:all_metric_plot}
\end{figure}

This study evaluated several similarity metrics $S(B_1,B_2)$ to compare two images $B_1$ and $B_2$. These include MIND, SSD, NMI, and NCC (Section \ref{subsec:metrics}), for OCT+LS alignment. Each metric was first evaluated using manually aligned registrations as ground truth (GT) and the corresponding similarity scores were recorded. The same metrics were then applied and the optimized similarity values were recorded and compared with the GT (Figure \ref{fig:metric_comparison}). Although some metrics achieved higher optimized scores during automatic registration, these values did not consistently correspond to visually correct alignment. 
Registration accuracy was further evaluated using mean target registration error (mTRE) (Figure \ref{fig:metric_plot}). NCC demonstrated the most consistent performance, achieving the lowest mTRE values among the evaluated metrics.
To further evaluate robustness, the same metrics were tested under constrained registration conditions across increasing translation bounds (0.5–2 mm) and rotation bounds ($5^\circ$--$40^\circ$). Figure \ref{fig:Constrained_metrics} shows the visual alignment results at the smallest search range, and Figure \ref{fig:all_metric_plot} shows the corresponding mTRE across all evaluated bounds. NCC maintained reliable visual alignment and achieved the lowest mean mTRE across all search ranges, demonstrating robustness to the choice of PSO bounds.
Therefore, NCC was selected as the primary similarity metric for OCT+LS alignment. NCC computes the correlation coefficient between corresponding intensity values of two images to quantify their similarity \cite{evangelidis2008parametric}:

\begin{equation}
\label{eqn:ncc}
    \mathrm{C}({B}_1,{B}_2)
    =\frac{\displaystyle\sum_{j=1}^{N}
    \bar{\mathbf{b}}_1[j] \, \bar{\mathbf{b}}_2[j]}
    {\displaystyle\sqrt{\sum_{j=1}^{N}
    \bar{\mathbf{b}}_1[j]^2}
    \;\sqrt{\sum_{j=1}^{N}
    \bar{\mathbf{b}}_2[j]^2}}
    =\frac{\bar{\mathbf{b}}_1 \cdot \bar{\mathbf{b}}_2}{
    \|\bar{\mathbf{b}}_1\| \, \|\bar{\mathbf{b}}_2\|}
\end{equation}

\subsection{Optimization}
\label{subsec:optimization}
\begin{figure}[H]
    \centering
    \includegraphics[width=1\linewidth]{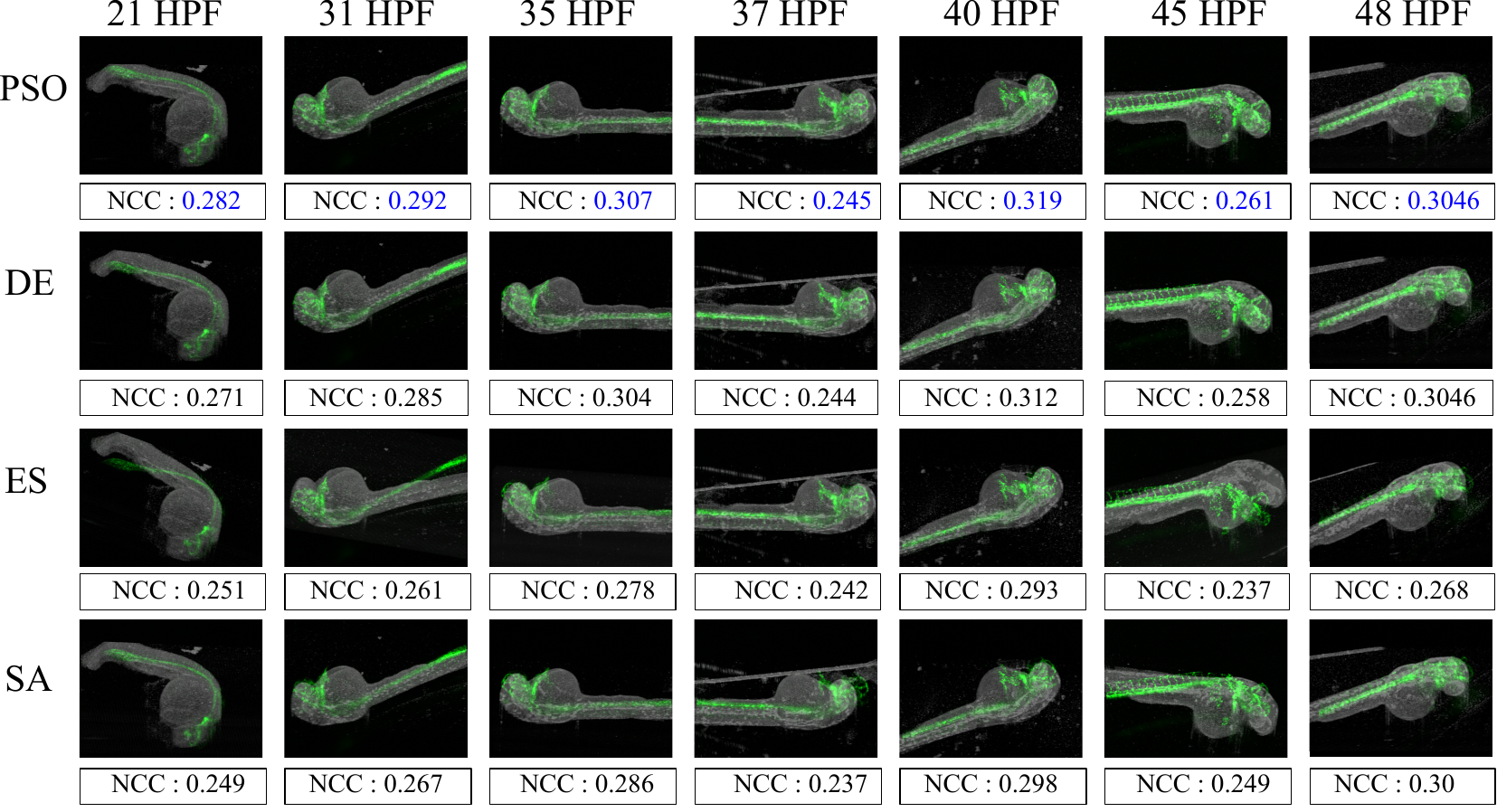}
    \caption{Comparison of optimization methods for aligning multimodal data with the NCC metric. Alignment results are shown for each time point along with the corresponding NCC metric values (higher is better). Blue values indicate the highest NCC value among the optimization methods at each time point.}
    \label{fig:optimization_comparison}
\end{figure}

\begin{figure}[H]
    \centering
    \includegraphics[width=0.75\linewidth]{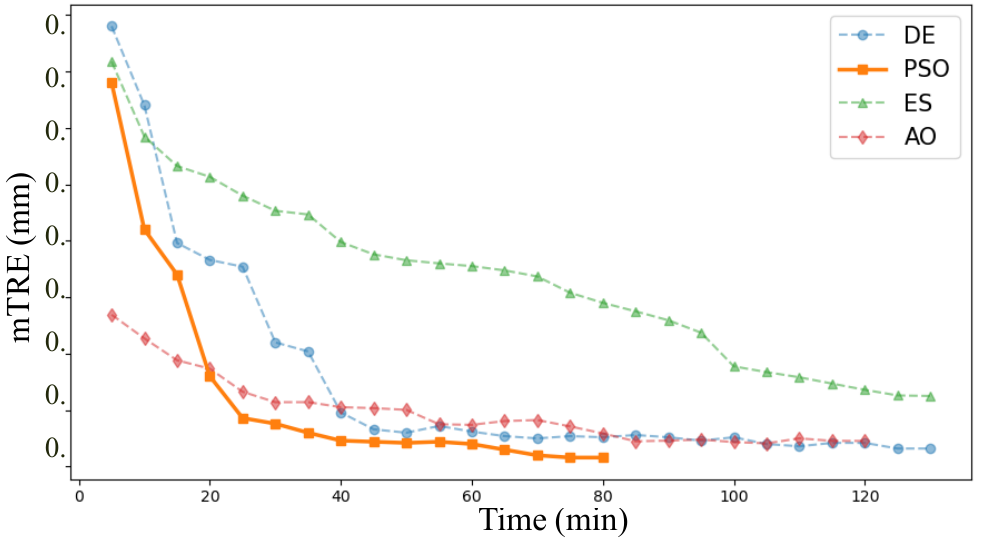}
    \caption{Comparison of registration convergence across optimization methods using mean target registration error (mTRE) over time. The plot shows the performance of DE, PSO, ES, and SA at each 5 minute intervals, where lower mTRE indicates better alignment.}
    \label{fig:optimization_plot}
\end{figure}

We tested four optimization algorithms for multimodal and temporal registration: Differential evolution (DE)\cite{das2010differential}, Simulated annealing (SA)\cite{guilmeau2021simulated}, ES \cite{li2020evolution}, and PSO. We found that PSO achieved higher NCC values compared with the other methods, indicating improved alignment quality (Figure \ref{fig:optimization_comparison}). Furthermore, the convergence of mTRE over time demonstrated that PSO converged faster and reached lower registration errors than the other optimization algorithms(Figure \ref{fig:optimization_plot}). Therefore, PSO was selected as the primary optimization method for the registration.

PSO is an iterative population-based method that maintains a set of $K$ particles. Each particle $n \in \{0, \ldots, K-1\}$ represents a candidate position in the search space $\bm{\psi}_n$ and a corresponding velocity $\mathbf{v}_n$ that determines its next search position. At each iteration $t \in \mathbb{N}$, all particles update their velocity and position based on three criteria:
\begin{itemize}
    \item the current velocity $\mathbf{v}_n$
    \item the best position $\mathbf{p}_n$ each particle identified (so far)
    \item the best global position $\mathbf{g}$ found by the entire swarm (so far)
\end{itemize}
Acceleration and velocity are updated using explicit integration \cite{ballerini2025particle}:
\begin{equation}
    \mathbf{a}_{n, t} =
    \left(
    \left[
        \mathbf{p}_n \ \ \mathbf{g}_{t-1}
    \right]
     - \bm{\psi}_{n,t-1}\mathbf{1}^T
     \right)
        \text{diag}(\mathbf{c})\mathbf{r}_{n,t}
    \label{eqn:pso_acceleration}
\end{equation}
\begin{equation}
    \mathbf{v}_{n,t} = w\,\mathbf{v}_{n,t-1} + \mathbf{a}_{n, t}
\label{eqn:pso_velocity}
\end{equation}
where $w$ is an inertia weight that controls how much the previous velocity affects the current velocity. $\mathbf{c}^T = [c_p \ c_g]$ are acceleration constants determining the influence of $\mathbf{p}_n$ and $\mathbf{g}_t$, and $\mathbf{r}_{n,t} \in [0,1]^2$ are random numbers between 0 and 1.
The position is updated as:
\begin{equation}
    \bm{\psi}_{n,t} = \bm{\psi}_{n,t-1} + \mathbf{v}_{n,t}
\label{eqn:pso_position}
\end{equation}

During the optimization, the NCC value is evaluated for all particles in the swarm every iteration, and PSO uses these values to update $\mathbf{p}_n$ and $\mathbf{g}_t$. Equations \ref{eqn:pso_acceleration}, \ref{eqn:pso_velocity}, and \ref{eqn:pso_position} are then used to determine the next iteration's samples. The process repeats until a convergence threshold is reached \cite{wachowiak2004approach, lin2012hybrid} or the PSO has exceeded a maximum number of iterations. In this study, the maximum number of iterations was set to 50.

\subsection{Registration pipeline}
The parameter vector for the optimization is given by $\bm{\psi}_{n,t} = [x,y,z, \theta_x, \theta_y, \theta_z]$, corresponding to all of the possible translations and rotations constrained by the OCT+LS imaging system. This requires two stages: (1) alignment of multimodal OCT and LS images at each acquisition time (\autoref{subsec:Pairwise}), and (2) alignment of the sample across all acquisition time points (\autoref{subsec:multitimesteps}). Both steps seek an optimal set of parameters $\bm{\psi}^*$ that maximize the similarity metric between target images:
\begin{equation}
    \bm{\psi}^* = \arg\max_{\bm{\psi}} \mathrm{C}\big(B_1, T_{\bm{\psi}}(B_2)\big)
\end{equation}

In the following, we detail the two stages of our registration pipeline. 

\subsection{Pairwise OCT+LS alignment}
\label{subsec:Pairwise}
We first perform a 3D rigid registration to align OCT and LS volumes to a common coordinate space within each sample time point. $L(\mathbf{u})$ denotes an LS volume of size $X_1 \times Y_1 \times Z_1$ and $O(\mathbf{v})$ denotes an OCT volume of size $X_2 \times Y_2 \times Z_2$, where $\mathbf{u}$ and $\mathbf{v}$ are voxel positions in the corresponding image spaces (indices). Both images represent an object in real space using coordinates $\mathbf{x} = [x,y,z,1]^T$. Aligning these images requires finding the relationship between real-space coordinates and voxel indices:
\begin{equation}
    \mathbf{u} = \mathbf{M}\mathbf{x}
    \quad \text{and} \quad
    \mathbf{v} = \mathbf{N}\mathbf{x}
\end{equation}

$\mathbf{x}$ is a position in real space (e.g., mm), $\mathbf{M}$ and $\mathbf{N}$ are $4 \times 4$ affine transformations that account for voxel scaling and are constrained to translation and Euler rotation. For example, given a specific spatial location $\mathbf{x}_i$, we calculate the corresponding voxel indices in $L(\mathbf{u}_i)$ and $O(\mathbf{v}_i)$ using:
\begin{equation}
    \mathbf{M}^{-1}\mathbf{u}_i = \mathbf{N}^{-1}\mathbf{v}_i = \mathbf{x}_i
\end{equation} 
 
\subsubsection{Step 1: Initialization}
We first calculate the initial matrices $\mathbf{N}_0$ and $\mathbf{M}_0$ that align the intensity centroids with the origin in world space and account for known voxel sizes for each modality. The matrix $\mathbf{M}_0 = \mathbf{T}_0 \mathbf{S}_0$ is composed of two transformations:
\begin{itemize}
    \item A (typically anisotropic) scaling transformation by a vector $\mathbf{s}_0$ that accounts for voxel spacing.
    \item A translation by a vector $\mathbf{t}_0$ that moves the centroid of each volume to the world-space origin.   
\end{itemize}

\begin{figure}[H]
    \centering
    \includegraphics[width=1\linewidth]{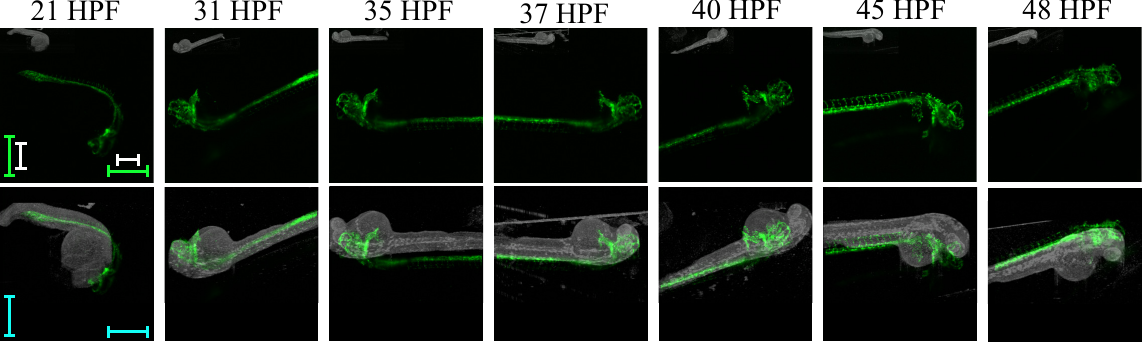}
	\caption{Correcting for voxel size and anisotropy. Sample rates differ between modalities and axes, so the first step is to compensate for these differences by calculating a scale transform $\mathbf{S}_0$ that maps both modalities into a common space. (top) initial images in pixel space are shown with scale bars representing 1 mm (green = LS, white = OCT). (bottom) After applying the scaling transform $\mathbf{S}_0$ (scale bars = 1 mm).}
    \label{fig:before_registration}
\end{figure}

The correction for voxel size and anisotropy during initialization is illustrated in Fig.~\ref{fig:before_registration}. Since the voxel sizes for each image are known, the optimization only accounts for translation and rotation. We fix the LS image $\mathbf{M} = \mathbf{M}_0$ at its acquired orientation and $\mathbf{N}$ is optimized from $\mathbf{N}_0$ to align the OCT to the LS volume. $\mathbf{N}$ is optimized over a world-space region defined by a 3D size $S_x \times S_y \times S_z \in \mathbb{R}^3$ (e.g., millimeters), with a corresponding sampling resolution of $N_x \times N_y \times N_z \in \mathbb{Z}^3$ voxels.

\subsubsection{Step 2: Optimization}
From the initial guesses $\mathbf{M}_0$ and $\mathbf{N}_0$, we iteratively optimize the parameters $[t_x, t_y, t_z,\theta_x, \theta_y, \theta_z]$ to compute progressively refined OCT transformations $\mathbf{N}_\psi$ while the LS transform remains fixed: $\mathbf{M}_n = \mathbf{M}_0$.

The optimization is performed using PSO (Equations~\ref{eqn:pso_velocity} and~\ref{eqn:pso_position}) to estimate parameters. Since PSO is formulated as a minimization procedure, we define the cost function as the negative of the NCC:
\begin{equation}
    \min_{\bm{\psi}=\{t_x, t_y, t_z, \theta_x, \theta_y, \theta_z\}} -\,\mathrm{C}\!\big[ L(\mathbf{M}\mathbf{x}),\; O(\mathbf{N}_\psi\mathbf{x}) \big]
\end{equation}
The transformation matrix $\mathbf{N}_\psi$ is composed of a rotation, translation, and fixed scaling with distinct parameters that are optimized:
\begin{equation}
    \mathbf{N}_\psi = \mathbf{T}(\mathbf{t}) \mathbf{R}(\theta_x, \theta_y, \theta_z) \mathbf{S}_0
\end{equation}

\subsection{Sequential registration between time points}
\label{subsec:multitimesteps}
\begin{figure}[H]
    \centering
    \includegraphics[width=1\linewidth]{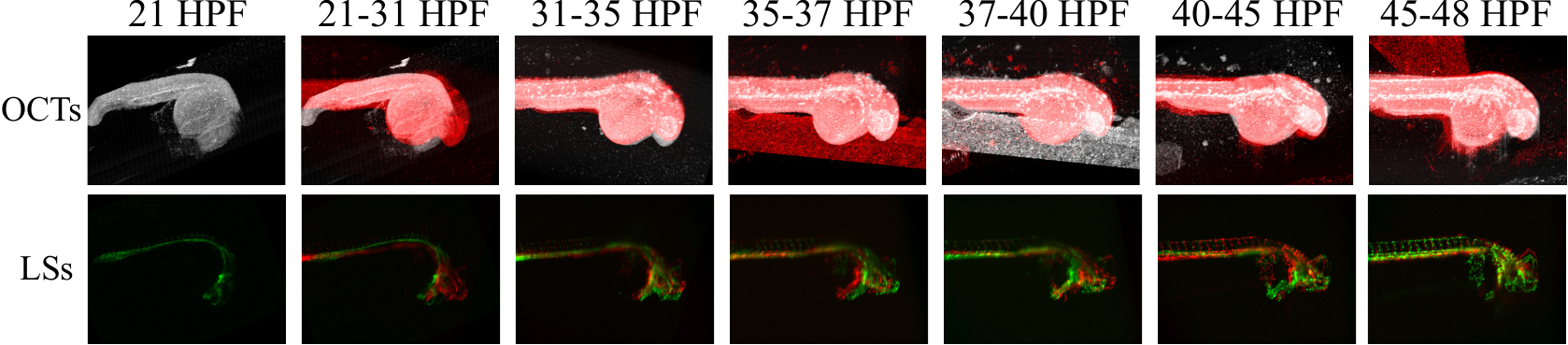}
    \caption{Sequential alignment results across timepoints (21 HPF - 48 HPF) for both OCT–OCT and LS–LS images using NCC similarity metric and PSO optimization.}
    \label{fig:temporal_registration1}
\end{figure}

After pairwise LS+OCT registration at each time point, temporal registration was performed between consecutive OCT volumes. OCT was selected for temporal registration because it provided consistent structural features across time points. The resulting transformation for each time point was subsequently applied to the corresponding LS volume (Figure \ref{fig:temporal_registration1}), preserving the spatial correspondence established during pairwise registration. Substantial changes in sample orientation and spatial position due to repositioning during imaging and embryo growth required a coarse prealignment step before PSO optimization (Section \ref{subsec:optimization}). The OCT volume acquired at time ${t+1}$ was subjected to candidate flips and rotations within the XZ plane, and the translation for each candidate was estimated using NCC. The candidate yielding the highest similarity to the OCT volume at time ${t}$ was used to initialize the registration.

This initial transformation was refined using the rigid registration framework described in Section (\ref{subsec:Pairwise}). Applying this procedure sequentially across consecutive time points mapped all OCT volumes into a common longitudinal coordinate system. Finally, the temporal transformations were transferred to the corresponding LS volumes using the pairwise OCT+LS relationships, preserving the multimodal alignment established at each time point.

\begin{figure}[H]
    \centering
    \includegraphics[width=1\linewidth]{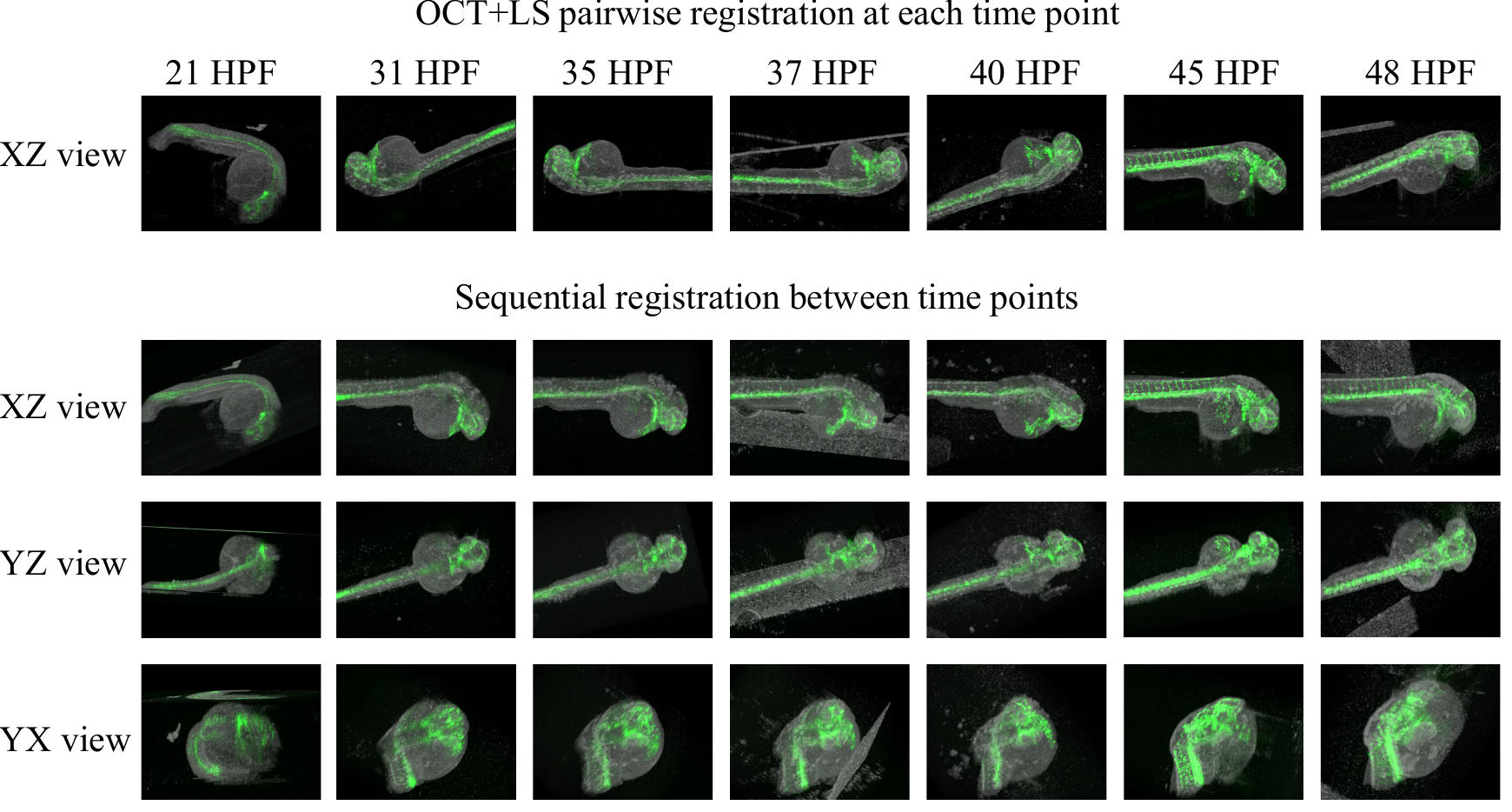}
    \caption{Registration results from 21 to 48 HPF. The first row shows pairwise OCT+LS registration at each time point. Following rows show sequential registration between time points in the YZ, ZX, and XY planes, respectively.}
    \label{fig:temporal_registration2}
\end{figure}

The longitudinal registration results are shown in Figure (\ref{fig:temporal_registration2}). The first row shows pairwise OCT+LS registration at each time point, followed by the sequential registration results in the XZ, YZ, and YX planes, demonstrating that multimodal alignment was maintained in the common longitudinal coordinate system.

\section{Conclusion}
The proposed framework provides a fully-automated approach for registering longitudinal OCT+LS images without manual landmark selection, a pre-existing segmentation, or feature correspondences between modalities. Unlike conventional feature-based methods, which can perform poorly when OCT and LS images contain different structural and molecular information, this approach uses image similarity and the physical constraints of the imaging system to achieve consistent alignment across modalities and developmental time points. The registered volumes allow tissue morphology captured by OCT to be directly compared with fluorescence information captured by LS. This enables longitudinal studies of how environmental exposures, such as alcohol and micro or nanoplastics \cite{huang2024exploring}, affect structural development and molecular expression in living zebrafish embryos.
Future work will extend this framework to additional transgenic zebrafish and fluorescent labels that highlight different cell types, proteins, and biological processes. Applying this method will allow the relationships between tissue morphology and a broader range of molecular processes to be investigated throughout development. Future studies may also incorporate deformable registration to account for localized anatomical changes that cannot be fully represented by rigid transformations.

The source code for the proposed implementation is available on GitHub (https://github.com/Mahshid-Babeli/octls-registration) with sample data sets available Zenodo \cite{babeli_nia_2026_23105189}.

\section*{Acknowledgments}
This work was funded in part by the National Institutes of Health grants \#R01HL146745 (KL, DM), \#R01HD086765 (KL), and \#R01AA028406 (KL) and the National Science Foundation grant \#1943455 (MN, DM).

\section*{Disclosures}
Kirill V. Larin and Manmohan Singh have a financial interest in ElastEye LLC., which is not directly related to this work.

\nolinenumbers

%
%
%

\bibliography{reference}

@misc{babeli_nia_2026_23105189,
  author       = {Babeli Nia, Mahshid},
  title        = {Zebrafish OCT and Light-Sheet Microscopy Dataset},
  month        = oct,
  year         = 2026,
  publisher    = {Zenodo},
  doi          = {10.5281/zenodo.23105189},
  url          = {https://doi.org/10.5281/zenodo.23105189},
}

@article{khajavi2021multimodal,
  title={Multimodal high-resolution embryonic imaging with light sheet fluorescence microscopy and optical coherence tomography},
  author={Khajavi, Behzad and Sun, Ruijiao and Chawla, Harshdeep Singh and Le, Henry H and Singh, Manmohan and Schill, Alexander W and Dickinson, Mary E and Mayerich, David and Larin, Kirill V},
  journal={Optics letters},
  volume={46},
  number={17},
  pages={4180--4183},
  year={2021},
  publisher={Optical Society of America}
}

@article{lawson2002vivo,
  title={In vivo imaging of embryonic vascular development using transgenic zebrafish},
  author={Lawson, Nathan D and Weinstein, Brant M},
  journal={Developmental biology},
  volume={248},
  number={2},
  pages={307--318},
  year={2002},
  publisher={Elsevier}
}

@article{keller2008reconstruction,
  title={Reconstruction of zebrafish early embryonic development by scanned light sheet microscopy},
  author={Keller, Philipp J and Schmidt, Annette D and Wittbrodt, Joachim and Stelzer, Ernst HK},
  journal={science},
  volume={322},
  number={5904},
  pages={1065--1069},
  year={2008},
  publisher={American Association for the Advancement of Science}
}

@article{vladimirov2014light,
  title={Light-sheet functional imaging in fictively behaving zebrafish},
  author={Vladimirov, Nikita and Mu, Yu and Kawashima, Takashi and Bennett, Davis V and Yang, Chao-Tsung and Looger, Loren L and Keller, Philipp J and Freeman, Jeremy and Ahrens, Misha B},
  journal={Nature methods},
  volume={11},
  number={9},
  pages={883--884},
  year={2014},
  publisher={Nature Publishing Group US New York}
}

@article{lichtenegger2022optical,
  title={Optical coherence tomography is a promising tool for zebrafish-based research—a review},
  author={Lichtenegger, Antonia and Baumann, Bernhard and Yasuno, Yoshiaki},
  journal={Bioengineering},
  volume={10},
  number={1},
  pages={5},
  year={2022},
  publisher={MDPI}
}

@article{heinrich2012mind,
  title={MIND: Modality independent neighbourhood descriptor for multi-modal deformable registration},
  author={Heinrich, Mattias P and Jenkinson, Mark and Bhushan, Manav and Matin, Tahreema and Gleeson, Fergus V and Brady, Michael and Schnabel, Julia A},
  journal={Medical image analysis},
  volume={16},
  number={7},
  pages={1423--1435},
  year={2012},
  publisher={Elsevier}
}

@article{evangelidis2008parametric,
  title={Parametric image alignment using enhanced correlation coefficient maximization},
  author={Evangelidis, Georgios D and Psarakis, Emmanouil Z},
  journal={IEEE transactions on pattern analysis and machine intelligence},
  volume={30},
  number={10},
  pages={1858--1865},
  year={2008},
  publisher={IEEE}
}

@inproceedings{kennedy1995particle,
  title={Particle swarm optimization},
  author={Kennedy, James and Eberhart, Russell},
  booktitle={Proceedings of ICNN'95-international conference on neural networks},
  volume={4},
  pages={1942--1948},
  year={1995},
  organization={ieee}
}

@article{lin2012hybrid,
  title={Hybrid particle swarm optimization and its application to multimodal 3D medical image registration},
  author={Lin, Chen-Lun and Mimori, Aya and Chen, Yen-Wei},
  journal={Computational intelligence and neuroscience},
  volume={2012},
  number={1},
  pages={561406},
  year={2012},
  publisher={Wiley Online Library}
}

@article{wachowiak2004approach,
  title={An approach to multimodal biomedical image registration utilizing particle swarm optimization},
  author={Wachowiak, Mark P and Smol{\'\i}kov{\'a}, Renata and Zheng, Yufeng and Zurada, Jacek M and Elmaghraby, Adel Said},
  journal={IEEE Transactions on evolutionary computation},
  volume={8},
  number={3},
  pages={289--301},
  year={2004},
  publisher={IEEE}
}

@article{ballerini2025particle,
  title={Particle Swarm Optimization in 3D Medical Image Registration: A Systematic Review: L. Ballerini},
  author={Ballerini, Lucia},
  journal={Archives of Computational Methods in Engineering},
  volume={32},
  number={1},
  pages={311--318},
  year={2025},
  publisher={Springer}
}

@article{zhang2025multi,
  title={Multi-resolution SAR and optical remote sensing image registration methods: A review, datasets, and future perspectives},
  author={Zhang, Wenfei and Zhao, Ruipeng and Yao, Yongxiang and Wan, Yi and Wu, Peihao and Li, Jiayuan and Li, Yansheng and Zhang, Yongjun},
  journal={arXiv preprint arXiv:2502.01002},
  year={2025}
}

@article{velesaca2024multimodal,
  title={Multimodal image registration techniques: a comprehensive survey},
  author={Velesaca, Henry O and Bastidas, Gisel and Rouhani, Mohammad and Sappa, Angel D},
  journal={Multimedia Tools and Applications},
  volume={83},
  number={23},
  pages={63919--63947},
  year={2024},
  publisher={Springer}
}

@article{lowe2004distinctive,
  title={Distinctive image features from scale-invariant keypoints},
  author={Lowe, David G},
  journal={International journal of computer vision},
  volume={60},
  number={2},
  pages={91--110},
  year={2004},
  publisher={Springer}
}

@inproceedings{rublee2011orb,
  title={ORB: An efficient alternative to SIFT or SURF},
  author={Rublee, Ethan and Rabaud, Vincent and Konolige, Kurt and Bradski, Gary},
  booktitle={2011 International conference on computer vision},
  pages={2564--2571},
  year={2011},
  organization={Ieee}
}

@article{ye2024fast,
  title={Fast and robust optical-to-SAR remote sensing image registration using region-aware phase descriptor},
  author={Ye, Yibin and Wang, Qinwei and Zhao, Hong and Teng, Xichao and Bian, Yijie and Li, Zhang},
  journal={IEEE Transactions on Geoscience and Remote Sensing},
  volume={62},
  pages={1--12},
  year={2024},
  publisher={IEEE}
}

@article{moghimi2021comparison,
  title={Comparison of keypoint detectors and descriptors for relative radiometric normalization of bitemporal remote sensing images},
  author={Moghimi, Armin and Celik, Turgay and Mohammadzadeh, Ali and Kusetogullari, Huseyin},
  journal={IEEE Journal of Selected Topics in Applied Earth Observations and Remote Sensing},
  volume={14},
  pages={4063--4073},
  year={2021},
  publisher={IEEE}
}

@article{lian2025csift,
  title={CSIFT: A Structural Feature Encoding Framework for Cross-Modal Image Registration},
  author={Lian, Lipeng and Chen, Leping and An, Daoxiang},
  journal={IEEE Journal of Selected Topics in Applied Earth Observations and Remote Sensing},
  year={2025},
  publisher={IEEE}
}

@article{zhang2025optical,
  title={Optical and SAR image registration based on feature constrained algorithm},
  author={Zhang, Wannan},
  journal={Scientific Reports},
  volume={15},
  number={1},
  pages={36373},
  year={2025},
  publisher={Nature Publishing Group UK London}
}

@article{li2019rift,
  title={RIFT: Multi-modal image matching based on radiation-variation insensitive feature transform},
  author={Li, Jiayuan and Hu, Qingwu and Ai, Mingyao},
  journal={IEEE Transactions on Image Processing},
  volume={29},
  pages={3296--3310},
  year={2019},
  publisher={IEEE}
}

@article{jonsson2022image,
  title={An image registration method for voxel-wise analysis of whole-body oncological PET-CT},
  author={J{\"o}nsson, Hanna and Ekstr{\"o}m, Simon and Strand, Robin and Pedersen, Mette A and Molin, Daniel and Ahlstr{\"o}m, H{\aa}kan and Kullberg, Joel},
  journal={Scientific Reports},
  volume={12},
  number={1},
  pages={18768},
  year={2022},
  publisher={Nature Publishing Group UK London}
}

@article{chen2025dual,
  title={A dual-domain framework for multimodal medical image registration: Optimizing phase consistency with LPC-GIMI},
  author={Chen, Shanshan and He, Kangjian and Xu, Dan and Shi, Hongzhen and Zeng, Rong},
  journal={Biomedical Signal Processing and Control},
  volume={99},
  pages={106809},
  year={2025},
  publisher={Elsevier}
}

@article{siebert2024convexadam,
  title={Convexadam: Self-configuring dual-optimization-based 3d multitask medical image registration},
  author={Siebert, Hanna and Gro{\ss}br{\"o}hmer, Christoph and Hansen, Lasse and Heinrich, Mattias P},
  journal={IEEE Transactions on Medical Imaging},
  volume={44},
  number={2},
  pages={738--748},
  year={2024},
  publisher={IEEE}
}

@article{walluscheck2023mr,
  title={MR-CT multi-atlas registration guided by fully automated brain structure segmentation with CNNs},
  author={Walluscheck, Sina and Canalini, Luca and Strohm, Hannah and Diekmann, Susanne and Klein, Jan and Heldmann, Stefan},
  journal={International journal of computer assisted radiology and surgery},
  volume={18},
  number={3},
  pages={483--491},
  year={2023},
  publisher={Springer}
}

@article{shojaei2025novel,
  title={Novel feature-based method for multi-modal biomedical image registration compared to intensity-based technique},
  author={Shojaei, Mohammad Javad and Yang, Lichen and Shojaei, Kazem and Doungchawee, Jeerapat and Vachet, Richard W},
  journal={Scientific Reports},
  volume={15},
  number={1},
  pages={28183},
  year={2025},
  publisher={Nature Publishing Group UK London}
}

@article{arora2024adaptive,
  title={An adaptive medical image registration using hybridization of teaching learning-based optimization with affine and speeded up robust features with projective transformation},
  author={Arora, Paluck and Mehta, Rajesh and Ahuja, Rohit},
  journal={Cluster Computing},
  volume={27},
  number={1},
  pages={607--627},
  year={2024},
  publisher={Springer}
}

@article{hering2022learn2reg,
  title={Learn2Reg: comprehensive multi-task medical image registration challenge, dataset and evaluation in the era of deep learning},
  author={Hering, Alessa and Hansen, Lasse and Mok, Tony CW and Chung, Albert CS and Siebert, Hanna and H{\"a}ger, Stephanie and Lange, Annkristin and Kuckertz, Sven and Heldmann, Stefan and Shao, Wei and others},
  journal={IEEE Transactions on Medical Imaging},
  volume={42},
  number={3},
  pages={697--712},
  year={2022},
  publisher={IEEE}
}

@article{honkamaa2023sitreg,
  title={SITReg: Multi-resolution architecture for symmetric, inverse consistent, and topology preserving image registration},
  author={Honkamaa, Joel and Marttinen, Pekka},
  journal={arXiv preprint arXiv:2303.10211},
  year={2023}
}

@article{krentzel2025clem,
  title={CLEM-Reg: an automated point cloud-based registration algorithm for volume correlative light and electron microscopy},
  author={Krentzel, Daniel and Elphick, Matou{\v{s}} and Domart, Marie-Charlotte and Peddie, Christopher J and Laine, Romain F and Shand, Cameron and Henriques, Ricardo and Collinson, Lucy M and Jones, Martin L},
  journal={Nature Methods},
  volume={22},
  number={9},
  pages={1923--1934},
  year={2025},
  publisher={Nature Publishing Group US New York}
}

@article{liu2026zero,
  title={Zero-shot Multi-Contrast Brain MRI Registration by Intensity Randomizing T1-weighted MRI (LUMIR25)},
  author={Liu, Hengjie and Dou, Yimeng and Xu, Di and Fu, Xinyi and Ruan, Dan and Sheng, Ke},
  journal={arXiv preprint arXiv:2602.06292},
  year={2026}
}

@article{chen2026beyond,
  title={Beyond the LUMIR challenge: The pathway to foundational registration models},
  author={Chen, Junyu and Wei, Shuwen and Honkamaa, Joel and Marttinen, Pekka and Zhang, Hang and Liu, Min and Zhou, Yichao and Tan, Zuopeng and Wang, Zhuoyuan and Wang, Yi and others},
  journal={Medical Image Analysis},
  pages={104175},
  year={2026},
  publisher={Elsevier}
}

@article{reithmeir2025model,
  title={From model based to learned regularization in medical image registration: A comprehensive review},
  author={Reithmeir, Anna and Spieker, Veronika and Sideri-Lampretsa, Vasiliki and Rueckert, Daniel and Schnabel, Julia A and Zimmer, Veronika A},
  journal={Medical Image Analysis},
  pages={103854},
  year={2025},
  publisher={Elsevier}
}

@article{andersson2007non,
  title={Non-linear registration, aka Spatial normalisation FMRIB technical report TR07JA2},
  author={Andersson, JL and Jenkinson, Mark and Smith, Stephen and others},
  journal={FMRIB Analysis Group of the University of Oxford},
  volume={2},
  number={1},
  pages={e21},
  year={2007}
}

@article{balakrishnan2019voxelmorph,
  title={Voxelmorph: a learning framework for deformable medical image registration},
  author={Balakrishnan, Guha and Zhao, Amy and Sabuncu, Mert R and Guttag, John and Dalca, Adrian V},
  journal={IEEE transactions on medical imaging},
  volume={38},
  number={8},
  pages={1788--1800},
  year={2019},
  publisher={IEEE}
}

@article{viola1997alignment,
  title={Alignment by maximization of mutual information},
  author={Viola, Paul and Wells III, William M},
  journal={International journal of computer vision},
  volume={24},
  number={2},
  pages={137--154},
  year={1997},
  publisher={Springer}
}

@incollection{lecun2002efficient,
  title={Efficient backprop},
  author={LeCun, Yann and Bottou, L{\'e}on and Orr, Genevieve B and M{\"u}ller, Klaus-Robert},
  booktitle={Neural networks: Tricks of the trade},
  pages={9--50},
  year={2002},
  publisher={Springer}
}

@article{ruder2016overview,
  title={An overview of gradient descent optimization algorithms},
  author={Ruder, Sebastian},
  journal={arXiv preprint arXiv:1609.04747},
  year={2016}
}

@article{kingma2014adam,
  title={Adam: A method for stochastic optimization},
  author={Kingma, Diederik P and Ba, Jimmy},
  journal={arXiv preprint arXiv:1412.6980},
  year={2014}
}

@article{rahimi2023review,
  title={A review on constraint handling techniques for population-based algorithms: from single-objective to multi-objective optimization},
  author={Rahimi, Iman and Gandomi, Amir H and Chen, Fang and Mezura-Montes, Efren},
  journal={Archives of computational methods in engineering},
  volume={30},
  number={3},
  pages={2181--2209},
  year={2023},
  publisher={Springer}
}

@article{shah2019multi,
  title={Multi-scale imaging and analysis identify pan-embryo cell dynamics of germlayer formation in zebrafish},
  author={Shah, Gopi and Thierbach, Konstantin and Schmid, Benjamin and Waschke, Johannes and Reade, Anna and Hlawitschka, Mario and Roeder, Ingo and Scherf, Nico and Huisken, Jan},
  journal={Nature communications},
  volume={10},
  number={1},
  pages={5753},
  year={2019},
  publisher={Nature Publishing Group UK London}
}

@article{dorent2025brain,
  title={The Brain Resection Multimodal Image Registration (ReMIND2Reg) 2025 Challenge},
  author={Dorent, Reuben and Rigolo, Laura and Galvin, Colin P and Chen, Junyu and Heinrich, Mattias P and Carass, Aaron and Colliot, Olivier and Wassermann, Demian and Golby, Alexandra and Kapur, Tina and others},
  journal={arXiv preprint arXiv:2508.09649},
  year={2025}
}

@article{juvekar2023brain,
  title={The brain resection multimodal imaging database (remind)},
  author={Juvekar, P and Dorent, R and K{\"o}gl, F and Torio, E and Barr, C and Rigolo, L and Galvin, C and Jowkar, N and Kazi, A and Haouchine, N},
  journal={The Cancer Imaging Archive},
  year={2023}
}

@article{das2010differential,
  title={Differential evolution: A survey of the state-of-the-art},
  author={Das, Swagatam and Suganthan, Ponnuthurai Nagaratnam},
  journal={IEEE transactions on evolutionary computation},
  volume={15},
  number={1},
  pages={4--31},
  year={2010},
  publisher={Ieee}
}

@article{li2020evolution,
  title={Evolution strategies for continuous optimization: A survey of the state-of-the-art},
  author={Li, Zhenhua and Lin, Xi and Zhang, Qingfu and Liu, Hailin},
  journal={Swarm and Evolutionary Computation},
  volume={56},
  pages={100694},
  year={2020},
  publisher={Elsevier}
}

@inproceedings{guilmeau2021simulated,
  title={Simulated annealing: A review and a new scheme},
  author={Guilmeau, Thomas and Chouzenoux, Emilie and Elvira, V{\'\i}ctor},
  booktitle={2021 IEEE statistical signal processing workshop (SSP)},
  pages={101--105},
  year={2021},
  organization={IEEE}
}

@article{shami2022particle,
  title={Particle swarm optimization: A comprehensive survey},
  author={Shami, Tareq M and El-Saleh, Ayman A and Alswaitti, Mohammed and Al-Tashi, Qasem and Summakieh, Mhd Amen and Mirjalili, Seyedali},
  journal={Ieee Access},
  volume={10},
  pages={10031--10061},
  year={2022},
  publisher={IEEE}
}

@article{huang2024exploring,
  title={Exploring developmental toxicity of microplastics and nanoplastics (MNPS): Insights from investigations using zebrafish embryos},
  author={Huang, Wenlong and Mo, Jiezhang and Li, Jiejie and Wu, Kusheng},
  journal={Science of the Total Environment},
  volume={933},
  pages={173012},
  year={2024},
  publisher={Elsevier}
}

@article{salimans2017evolution,
  title={Evolution strategies as a scalable alternative to reinforcement learning},
  author={Salimans, Tim and Ho, Jonathan and Chen, Xi and Sidor, Szymon and Sutskever, Ilya},
  journal={arXiv preprint arXiv:1703.03864},
  year={2017}
}

\end{document}